# ADAPTIVE AND DYNAMIC WIRELESS ROUTERS WITH SMART ANTENNA FOR POWER MANAGEMENT

S.Venkata Krishnan, R.Sriram and N.Senthil Kumar

Department of Computer Science Engineering, Sri Sai Ram Engineering College,
Chennai, Tamil Nadu, India
anirudhvenkats@yahoo.co.in
r.sriram.18@gmail.com
senthilneels@gmail.com

*ABSTRACT*

*In the recent evolution of wireless technologies, the power management has been a worrying factor. In order to overcome the power shortage, steps are taken to find new kind of energy harvesting methods, power attenuation reduction methods and power saving techniques. Wireless routers even though consume not much of power, battery powered devices require a lot. Omni directional antenna embedded with multiple antennae focusing the beam of radio wave signals in the direction of nodes with least transmission angle can be a solution for this problem which is called as "Smart Antenna". To reduce power maceration we are going for adaptive and dynamic transmission wherein the transmission angle of antennae is varied in accordance with the movement of nodes. Apart from saving the power considerably, it also improves the signal strength*

## 1. INTRODUCTION

The world is completely moving towards wireless technologies especially Bluetooth, Wi-Fi, WiMAX and many other new technologies. Data transmission is the inevitable part of wireless communication and effective means for the same has been a major concern. In these wireless technologies power losses has been playing a major role. The problem of power loss due to the transmission of signals in the regions where there is no node needs to be addressed. Wi-Fi routers are commonly used these days. Though it has some advantages of replacing complex wires, portability and also data rates competing with that of wired networks like Ethernets, it carries the disadvantages such as power loss due to interference, low signal strength and low range.

Wi-Fi uses two kinds of antenna:
1) Omni directional antenna
2) Hi-gain directional antenna

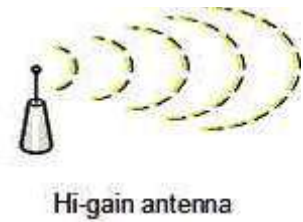

Figure 1.Hi-gain Antenna

Hi-gain directional antenna limits the range of the router to a confined region, so to avoid this constraint smart antenna having multiple antennae embedded within a single antenna is used. Smart antenna which is an array of antennas, changes the phases of the subsequent antennas relatively so as to avoid any kind of interferences is called as Phased array antenna [2].

Smart Antennas are of 2 types namely,
1. Switched Beam Antenna
2. Adaptive Array Antenna

In this technique, Adaptive array antennas are going to be used to service the nodes at different loads [1][10]. The switched beam antennas which form a fixed beam radiation pattern according to the loads can also be used. But by using adaptive array the power wastages that occur in switched beam can be reduced. The antenna will adaptively and dynamically change its angle of transmission of radio waves according to the movement of the mobile nodes which is assigned to service the particular node [7]. One antenna will intermittently broadcast its unique name for new nodes in all directions. At any given instance, angle between the nodes's starting point and ending point will be calculated and the least angle will be taken for transmission. By this way one can reduce the power loss in situations where there is only one node or minimum number of nodes in the service range. Using directional nature of the antenna alone does not serve the purpose instead it will reduce the range of the router.

## 2. IMPLEMENTATION

The main idea in this proposal is to reduce the power consumption of wireless routers which are Omni directional in nature. Wireless routers, macerate most of their power in transmitting radio wave signals with their *S*hort *f*or *S*ervice *S*et *I*dentifier, a 32 bit character unique identifier for the nodes to connect to that BSS is often abbreviated as SSID. If the number of nodes serviced is less, then there is a huge power loss.

So in order to reduce this power wastage we are going in for multiple antennae embedded within a single antenna. Adaptive array of antennas whose beam doesn't overlap with each other can be used; it has the added advantage that it allows two users to use the same antenna simultaneously. For example, the maximum number of nodes which a router can service at any given time is say 256 clients. If we assume that there is only 1 node then the angle of transmission is confined to that node only. If the transmission is carried for only one node, then there will be needless wastage of power because no other node is receiving.

So the array of antennas embedded in a single antenna with one antenna broadcasting its SSID Omni directionally with a less intense beam than the normal and will consume very low power at that time and it also sends signals intermittently.

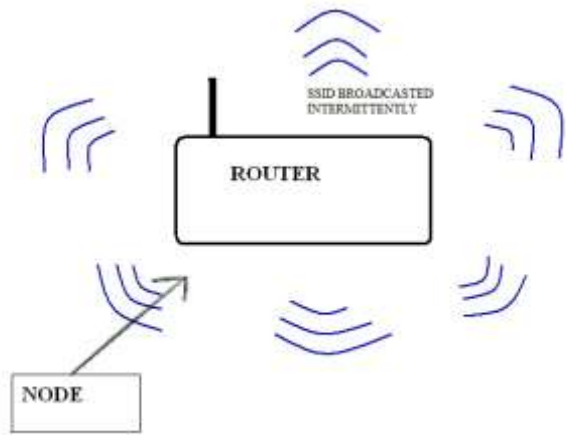

Figure 2.Advent of new node

Adaptive array carries another feature. Spatial diversity of the adaptive array antenna is the concept where the waves of signal are dense in the required direction and less in other direction. It can even nullify the signal waves in the direction where there is no load.

The three scenarios potraying how the transmission will be are:

## 2.1. Case (I):

A new node comes for service to the nearest Access Point and is stationed at a place: At that time, if a new node comes and tries to attach to it, adaptive nature of the antenna calculates the direction of arrival (DOA) and tunes the antenna into that frequency of that node. It sends the beam pattern to that node in that direction using the phenomenon called Beam forming.

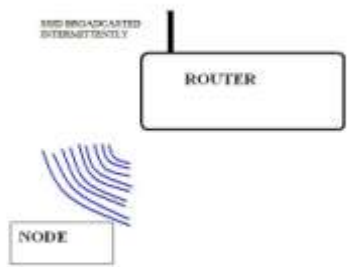

Figure 3.servicing of new node

Beam forming is the process of adding the phases of the signals constructively in the target's direction thereby nullifying the signals in the undesired direction [3]. By making the angle of transmission of the antennae small and converging to the serviceable node only, the signal strength of the node will be greater and the power consumption of the router will also be reduced in large quantum.This also fixes the problem of packet loss due to low signal strength.

Direction of arrival of the request from the nodes can be sensed by using techniques like MUSIC (Multiple Signal classification), ESPRIT (Estimation of signal parameters via rotational invariance techniques) and Matrix pencil method. Matrix pencil method is the most used and efficient method is one of the derivatives of the former 2 techniques.

## 2.2. Case (II):

When maximum nodes are getting serviced SSID will not be broadcasted: If maximum number of clients are getting serviced, the adaptive nature of the antenna will have its beam pattern towards the nodes which needs service and it will dynamically alter its signal power according to the frequency of requests in order to maximize the throughput of the routers. And the most important part is that it will not take any more requests. This is accomplished by not broadcasting SSID and later it starts broadcasting its SSID only if some node detaches its connection. Usually in wireless routers the maximum number of nodes that can get service at a time is 128 or 256.

But this idea is to reduce power consumption during worst cases where in the number of nodes to be serviced is minimal or it is not substantial. So this adaptive and dynamic transmission reduces the needless power consumption in meager load scenarios.

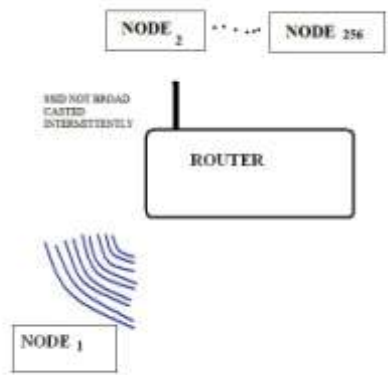

Figure 4.256 nodes getting serviced all at a time

## 2.3. Case (III):

When the node is mobile in nature: Once a node moves away from the BSS, the antenna which serviced that node will again start broadcasting its SSID intermittently. The adaptive nature of the antenna will be continuously monitoring the movement of the mobile node and accordingly will change its angle of transmission with respect to the access point [5]. Once it moves away from the range of the access point it will no longer get services from the access point.

The angle between the moving node and the access point is calculated as $\Theta_a$ called as azimuth angle of direction of arrival. The calculated angle is evaluated in such a way that it is minimum.

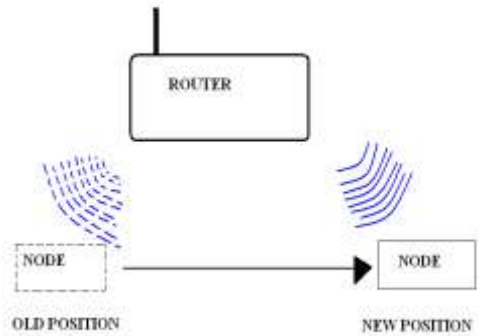

Figure 5.Mobile node getting adaptively serviced by the router

## 3. PERFORMANCE CHARACTERISTICS

Performance of this kind of systems is evaluated using a technique called Cross correlation. The envelope of cross correlation is a measure of quantity of performance. If we take to signals ($x_1(t)$ and $x_2(t)$), it will be processed using dual diversity system to create an improved signal $x_c(t)$.

The average signal strength at each antenna can be calculated as:
$P_1 = E\ (mod(x_1(t))^2)$
$P_2 = E\ (mod(x_2(t))^2)$

Statistical Expectation of the two signal strength of the antenna will give cross correlation of the signals from the two antennas.

For better performance then,

- Envelope cross correlation $\rho_e$ should be approximately equal to the square of the absolute value of $\rho_c$.
- Good diversity gain is possible when $\rho_e < 0.5$.

## 4. FUTURE WORKS

Though this idea includes multiple antennas at the transmission end to reduce the power consumption in places where the load is high, it is also used to increase the signal strength and range of that access point. But it can be used in both transmission and receiver end so as to increase the signal strength even better, the loss of packets due to fluctuations in power can be handled and also the basic security concerns of tapping signals by unwanted sources. MIMO (Multiple inputs and Multiple outputs) [4] is used in WiMAX in recent times and it will also be implemented soon in Wi-Fi to have better throughput and also some research works are going in these areas.

## 5. CONCLUSION

Our proposal provides a better solution for reducing the power loss produced by the wireless routers without compromising its range, security and its vital parameters. By using the array of antenna using adaptive array concept helps in having multiple antenna in a single structure as well as it can adaptively modify its nature. This helps in reducing needless wastage of power. Implementation of this system will increase the merits of the wireless communication to a great extent by increasing the range, signal strength and also better throughput by reducing power loss, packet loss and also security concerns. If the efficiency of the data transmission needs to be increased in a wireless system, then the signal strength, range of the router servicing the node and the power loss needs to be effectively managed. Hence our proposal provides a noteworthy solution for reduction in power loss and better signal strength in IEEE 802.11 family of systems.

## ACKNOWLEDGEMENTS

We want to thank all our faculty members who supported us and gave their valuable inputs. We also want to thank our family for their moral support to complete our idea.

## REFERENCES

[1]   "Adaptive Array antenna (Patent)", Robert Milne Patent number: 4700197 Issue date: 13 Oct 1987

[2]   T Do-Hong, P Russer, "Signal Processing For Wideband Smart Antenna Array Applications" IEEE Microwave Magazine (2004) Volume: 5, Issue: 1, Publisher: IEEE, Pages: 57-67

[3]   Viswanath, P.;   Tse, D.N.C.;   Laroia, R.; "Opportunistic Beam forming using Dumb Antennas" IEEE transactions on Information Theory, vol. 48, no. 6, June 2002 Page(s): 1277 – 1294

[4].   Ms.AR.Rajini, and M.M.Kanthimathi "Adaptive MIMO-OFDM Scheme with reduced computational Complexity & improved Capacity" International Journal of Computer Science and Information Security.

[5]   C.K. KO and R.D. Murch, A Diversity antenna for external mounting on wireless handsets, IEEE transactions on antennas and propagation, Vol 49, no 5, may 2001

[6]   Mobile Power Management for Wireless Communication Networks by John M. Rulnick and Nicholas Bambos

[7]   Dynamic Power Management in Wireless Sensor Networks by Amit Sinha and Anantha Chandrasekaran

[8]   Frank B. Gross, "Smart Antennas for Wireless Communications with Matlab", McGraw-Hill, 2005

[9]   Nathan Blaunstein, Christos G. Christodoulou "Radio Propagation and adaptive antenna for wireless communication "

[10]   Widrow, B.;   Mantey, P.E.;   Griffiths, L.J.;   Goode, B.B.  Adaptive Antenna Systems Page 2143– 2159.

Authors

S.Venkata Krishnan

R.Sriram

N.Senthil Kumar

Short Biography

We are final year students pursuing Computer Science and engineering in the esteemed institution Sri Sai Ram Engineering College, Affiliated to Anna University.